\documentclass[aps,prl,twocolumn,showpacs,superscriptaddress,groupedaddress]{revtex4}  
\usepackage{graphicx}  
\usepackage{dcolumn}   
\usepackage{bm}        
\usepackage{amssymb,amsmath}   

\usepackage{epsfig}
\usepackage{psfrag}
\usepackage{xspace}
\usepackage{multirow}
\usepackage{lscape}
\usepackage{comment}

\newcommand{\sigmatt}{\mbox{${\sigma}_{t\bar{t}}$}\xspace}

\newcommand{\dzero}     {D0\xspace}
\newcommand{\ttbar}     {\mbox{$t\bar{t}$}\xspace}
\newcommand{\tbar}      {\mbox{$\bar{t}$}\xspace}
\newcommand{\ppbar}     {\mbox{$p\bar{p}$}\xspace}
\newcommand{\qqbar}     {\mbox{$q\bar{q}$}\xspace}
\newcommand{\eplus}	{\mbox{$e$+jets}\xspace}
\newcommand{\muplus}    {\mbox{$\mu$+jets}\xspace}
\newcommand{\lplus}     {\mbox{$\ell$+jets}\xspace}
\newcommand{\ejets}	{\mbox{$e$+jets}\xspace}
\newcommand{\mujets}    {\mbox{$\mu$+jets}\xspace}
\newcommand{\ljets}     {\mbox{$\ell$+jets}\xspace}
\newcommand{\pythia}    {\mbox{\textsc{pythia}}\xspace}
\newcommand{\alpgen}    {\mbox{\textsc{alpgen}}\xspace}
\newcommand{\mcatnlo}   {\mbox{\textsc{mc@nlo}}\xspace}
\newcommand{\geant}     {\mbox{\textsc{geant}}\xspace}
\newcommand{\lllumi}    {5.4~fb$^{-1}$\xspace}
\newcommand{\ljetslumi} {5.3~fb$^{-1}$\xspace}

\newcommand{\GeV}{\ensuremath{\mathrm{Ge\kern-0.1em V}}\xspace}

\newcommand{\met}{\mbox{\ensuremath{p\kern-0.4em\slash_T}}\xspace}
\newcommand{\metx}{\mbox{\ensuremath{p\kern-0.6em\slash_x}}\xspace}
\newcommand{\mety}{\mbox{\ensuremath{p\kern-0.6em\slash_y}}\xspace}

\newcommand{\asm}{\mbox{$A_{\rm SM}$}\xspace}
\newcommand{\fmeas}{\mbox{$f_{\text {meas}}$}\xspace}

\newcommand{\ameas}{\mbox{$A_{\text {meas}}$}\xspace}

\newcommand{\herwig}    {\textsc{herwig}\xspace}

\newcommand{\ttb}{\mbox{$t\bar{t}$}\xspace}

\begin{document}

\hspace{5.2in} \mbox{Fermilab-Pub-11/555-E}

\title{Evidence for spin correlation in \boldmath \ttbar production}
%
\affiliation{Universidad de Buenos Aires, Buenos Aires, Argentina}
\affiliation{LAFEX, Centro Brasileiro de Pesquisas F{\'\i}sicas, Rio de Janeiro, Brazil}
\affiliation{Universidade do Estado do Rio de Janeiro, Rio de Janeiro, Brazil}
\affiliation{Universidade Federal do ABC, Santo Andr\'e, Brazil}
\affiliation{Instituto de F\'{\i}sica Te\'orica, Universidade Estadual Paulista, S\~ao Paulo, Brazil}
\affiliation{University of Science and Technology of China, Hefei, People's Republic of China}
\affiliation{Universidad de los Andes, Bogot\'{a}, Colombia}
\affiliation{Charles University, Faculty of Mathematics and Physics, Center for Particle Physics, Prague, Czech Republic}
\affiliation{Czech Technical University in Prague, Prague, Czech Republic}
\affiliation{Center for Particle Physics, Institute of Physics, Academy of Sciences of the Czech Republic, Prague, Czech Republic}
\affiliation{Universidad San Francisco de Quito, Quito, Ecuador}
\affiliation{LPC, Universit\'e Blaise Pascal, CNRS/IN2P3, Clermont, France}
\affiliation{LPSC, Universit\'e Joseph Fourier Grenoble 1, CNRS/IN2P3, Institut National Polytechnique de Grenoble, Grenoble, France}
\affiliation{CPPM, Aix-Marseille Universit\'e, CNRS/IN2P3, Marseille, France}
\affiliation{LAL, Universit\'e Paris-Sud, CNRS/IN2P3, Orsay, France}
\affiliation{LPNHE, Universit\'es Paris VI and VII, CNRS/IN2P3, Paris, France}
\affiliation{CEA, Irfu, SPP, Saclay, France}
\affiliation{IPHC, Universit\'e de Strasbourg, CNRS/IN2P3, Strasbourg, France}
\affiliation{IPNL, Universit\'e Lyon 1, CNRS/IN2P3, Villeurbanne, France and Universit\'e de Lyon, Lyon, France}
\affiliation{III. Physikalisches Institut A, RWTH Aachen University, Aachen, Germany}
\affiliation{Physikalisches Institut, Universit{\"a}t Freiburg, Freiburg, Germany}
\affiliation{II. Physikalisches Institut, Georg-August-Universit{\"a}t G\"ottingen, G\"ottingen, Germany}
\affiliation{Institut f{\"u}r Physik, Universit{\"a}t Mainz, Mainz, Germany}
\affiliation{Ludwig-Maximilians-Universit{\"a}t M{\"u}nchen, M{\"u}nchen, Germany}
\affiliation{Fachbereich Physik, Bergische Universit{\"a}t Wuppertal, Wuppertal, Germany}
\affiliation{Panjab University, Chandigarh, India}
\affiliation{Delhi University, Delhi, India}
\affiliation{Tata Institute of Fundamental Research, Mumbai, India}
\affiliation{University College Dublin, Dublin, Ireland}
\affiliation{Korea Detector Laboratory, Korea University, Seoul, Korea}
\affiliation{CINVESTAV, Mexico City, Mexico}
\affiliation{Nikhef, Science Park, Amsterdam, the Netherlands}
\affiliation{Radboud University Nijmegen, Nijmegen, the Netherlands and Nikhef, Science Park, Amsterdam, the Netherlands}
\affiliation{Joint Institute for Nuclear Research, Dubna, Russia}
\affiliation{Institute for Theoretical and Experimental Physics, Moscow, Russia}
\affiliation{Moscow State University, Moscow, Russia}
\affiliation{Institute for High Energy Physics, Protvino, Russia}
\affiliation{Petersburg Nuclear Physics Institute, St. Petersburg, Russia}
\affiliation{Instituci\'{o} Catalana de Recerca i Estudis Avan\c{c}ats (ICREA) and Institut de F\'{i}sica d'Altes Energies (IFAE), Barcelona, Spain}
\affiliation{Stockholm University, Stockholm and Uppsala University, Uppsala, Sweden}
\affiliation{Lancaster University, Lancaster LA1 4YB, United Kingdom}
\affiliation{Imperial College London, London SW7 2AZ, United Kingdom}
\affiliation{The University of Manchester, Manchester M13 9PL, United Kingdom}
\affiliation{University of Arizona, Tucson, Arizona 85721, USA}
\affiliation{University of California Riverside, Riverside, California 92521, USA}
\affiliation{Florida State University, Tallahassee, Florida 32306, USA}
\affiliation{Fermi National Accelerator Laboratory, Batavia, Illinois 60510, USA}
\affiliation{University of Illinois at Chicago, Chicago, Illinois 60607, USA}
\affiliation{Northern Illinois University, DeKalb, Illinois 60115, USA}
\affiliation{Northwestern University, Evanston, Illinois 60208, USA}
\affiliation{Indiana University, Bloomington, Indiana 47405, USA}
\affiliation{Purdue University Calumet, Hammond, Indiana 46323, USA}
\affiliation{University of Notre Dame, Notre Dame, Indiana 46556, USA}
\affiliation{Iowa State University, Ames, Iowa 50011, USA}
\affiliation{University of Kansas, Lawrence, Kansas 66045, USA}
\affiliation{Kansas State University, Manhattan, Kansas 66506, USA}
\affiliation{Louisiana Tech University, Ruston, Louisiana 71272, USA}
\affiliation{Boston University, Boston, Massachusetts 02215, USA}
\affiliation{Northeastern University, Boston, Massachusetts 02115, USA}
\affiliation{University of Michigan, Ann Arbor, Michigan 48109, USA}
\affiliation{Michigan State University, East Lansing, Michigan 48824, USA}
\affiliation{University of Mississippi, University, Mississippi 38677, USA}
\affiliation{University of Nebraska, Lincoln, Nebraska 68588, USA}
\affiliation{Rutgers University, Piscataway, New Jersey 08855, USA}
\affiliation{Princeton University, Princeton, New Jersey 08544, USA}
\affiliation{State University of New York, Buffalo, New York 14260, USA}
\affiliation{Columbia University, New York, New York 10027, USA}
\affiliation{University of Rochester, Rochester, New York 14627, USA}
\affiliation{State University of New York, Stony Brook, New York 11794, USA}
\affiliation{Brookhaven National Laboratory, Upton, New York 11973, USA}
\affiliation{Langston University, Langston, Oklahoma 73050, USA}
\affiliation{University of Oklahoma, Norman, Oklahoma 73019, USA}
\affiliation{Oklahoma State University, Stillwater, Oklahoma 74078, USA}
\affiliation{Brown University, Providence, Rhode Island 02912, USA}
\affiliation{University of Texas, Arlington, Texas 76019, USA}
\affiliation{Southern Methodist University, Dallas, Texas 75275, USA}
\affiliation{Rice University, Houston, Texas 77005, USA}
\affiliation{University of Virginia, Charlottesville, Virginia 22901, USA}
\affiliation{University of Washington, Seattle, Washington 98195, USA}
\author{V.M.~Abazov} \affiliation{Joint Institute for Nuclear Research, Dubna, Russia}
\author{B.~Abbott} \affiliation{University of Oklahoma, Norman, Oklahoma 73019, USA}
\author{B.S.~Acharya} \affiliation{Tata Institute of Fundamental Research, Mumbai, India}
\author{M.~Adams} \affiliation{University of Illinois at Chicago, Chicago, Illinois 60607, USA}
\author{T.~Adams} \affiliation{Florida State University, Tallahassee, Florida 32306, USA}
\author{G.D.~Alexeev} \affiliation{Joint Institute for Nuclear Research, Dubna, Russia}
\author{G.~Alkhazov} \affiliation{Petersburg Nuclear Physics Institute, St. Petersburg, Russia}
\author{A.~Alton$^{a}$} \affiliation{University of Michigan, Ann Arbor, Michigan 48109, USA}
\author{G.~Alverson} \affiliation{Northeastern University, Boston, Massachusetts 02115, USA}
\author{G.A.~Alves} \affiliation{LAFEX, Centro Brasileiro de Pesquisas F{\'\i}sicas, Rio de Janeiro, Brazil}
% ...
%
%
\collaboration{The D0 Collaboration\footnote{with visitors from
$^{a}$Augustana College, Sioux Falls, SD, USA,
$^{b}$The University of Liverpool, Liverpool, UK,
$^{c}$UPIITA-IPN, Mexico City, Mexico,
$^{d}$SLAC, Menlo Park, CA, USA,
$^{e}$University College London, London, UK,
$^{f}$Centro de Investigacion en Computacion - IPN, Mexico City, Mexico,
$^{g}$ECFM, Universidad Autonoma de Sinaloa, Culiac\'an, Mexico,
and 
$^{h}$Universit{\"a}t Bern, Bern, Switzerland.
$^{\ddag}$Deceased.
}} \noaffiliation
\vskip 0.25cm
\date{October 18, 2011}

\begin{abstract}
We present a measurement of the ratio of events with correlated
$t$ and $\bar{t}$ spins to the total number of \ttbar events.
This ratio $f$ is evaluated using a matrix-element-based approach in 
729 \ttbar candidate events with a single lepton $\ell$ (electron or muon) 
and at least four jets. The analyzed $p\bar{p}$ collisions data 
correspond to an integrated luminosity of \ljetslumi and were 
collected with the \dzero detector at the Fermilab Tevatron collider 
operating at a center-of-mass energy $\sqrt{s}=1.96$~TeV. 
Combining this result with a recent measurement of $f$ in 
dileptonic final states, we find $f$ in agreement with the standard model.
In addition, the combination provides evidence for the presence of  
spin correlation in \ttbar events with a significance of more
than 3 standard deviations. 
\end{abstract}

\maketitle



%

Although top and antitop quarks are produced unpolarized at hadron
colliders, their spin correlation can be measured, and a significant correlation 
is expected in the standard model (SM). The strength of
spin correlation depends on the production mechanism and differs, for
example, for \qqbar and $gg$ induced
\ttbar\ production~\cite{Mahlon_Parke}.  Since the top quark decays 
through the  electroweak interaction before it can interact through the 
strong interaction~\cite{Bigi:1986jk,d0_width},  the spin orientation of the top
quark at production is reflected  in the angular distributions of the
final state particles~\cite{spin_theory}. 
We present a measurement of the spin correlation of the $t$
and \tbar quarks to check its consistency with that expected in the
SM. 

The \ttbar\ spin correlation strength $C$ is defined by
${\rm d}^{2}\sigma_{t\bar{t}} /({\rm d}\cos\theta_1 {\rm d}\cos\theta_2)
=\sigma_{t\bar{t}} ( 1 - C \cos\theta_1 \cos\theta_2 )/4 $,
where $\sigma_{t\bar{t}}$ denotes the \ttbar production cross section,
and $\theta_1$, 
$\theta_2$ the angles between the spin-quantization axis and the
direction of flight of the down-type fermion from the $W$ boson decay
in the respective parent $t$ or \tbar rest frame. 
It is related to the fractional difference $A=(N_a-N_o)/(N_a+N_o)$
in the number of events $N_a$ where the top and antitop quark spins
are aligned and those where the top quarks spins have opposite alignment, $N_o$,
by $C= A |\alpha_1 \alpha_2|$ where $\alpha_i$ is the
spin analyzing power of the final state fermion under consideration.
In next-to-leading-order quantum chromodynamics (NLO QCD) $\alpha_{\ell^+}=1$ 
for the charged lepton in $t \rightarrow \ell^+ \nu_\ell b$ decays
and $\alpha_{\bar{d}}=0.97$ for the antidown quark in $t \rightarrow
\bar{d} u b$ decays~\cite{pol_power}. 
The value $A=+1$ ($-1$) corresponds to fully parallel (antiparallel) 
spins. Using the beam momentum vector as the quantization axis, the SM
predicts $\asm=0.78^{+0.03}_{-0.04}$ at NLO QCD
for \ppbar collisions at $\sqrt{s}=1.96$ TeV~\cite{spin_theory}. 

Three \ttbar spin correlation measurements based on the double
differential angular distribution 
have been published so far~\cite{D0RunI,cdf_spin,d0_spin}.
However, none of them had sufficient sensitivity to
distinguish between the hypothesis of spin correlation, as predicted by
the SM, and no spin correlation. A fourth measurement  was performed by the 
\dzero Collaboration in an analysis of 
\lllumi of integrated luminosity in the \ttbar dilepton channel, and
reached an expected 
sensitivity  of 3 standard deviations (SDs) from the no-correlation 
hypothesis. In that
analysis~\cite{d0_spin_me}, leading-order (LO) matrix elements (MEs) were 
used to measure the ratio $f$ of events with correlated $t$ and \tbar 
spins to the total number of \ttbar events by comparing Monte Carlo (MC) 
simulations with SM spin correlation and without spin correlation to data.

In this Letter, we present the first measurement of the ratio $f$ 
using the matrix-element approach in \ttbar \lplus
events. 
The $t$ and \tbar quarks are each assumed to decay into a $W$ boson and 
a $b$ quark, with one of the $W$ bosons decaying directly or  
via a leptonic tau decay into an electron or muon and the
corresponding neutrinos and the other $W$ boson decaying into two
quarks. 
We use \ljetslumi of integrated luminosity collected with the 
\dzero detector at the Fermilab Tevatron \ppbar collider and
combine our results with the corresponding measurement in the dilepton
channel~\cite{d0_spin_me}. 

%
%
A description of the \dzero detector can be found elsewhere~\cite{run2det}.  
%
%
%
%
%
We use the same event selections as in the measurement of \sigmatt\ in the 
\lplus channel~\cite{xsec_paper}. We require 
one isolated electron with transverse momentum  
$p_T>20$~\GeV and pseudorapidity $|\eta|<1.1$~\cite{eta}, 
or one isolated muon with $p_T>20$~\GeV and $|\eta|<2.0$, 
as well as an imbalance in transverse momentum $\met>20~(25)~\GeV$ for
the \ejets\ (\mujets) channel. Events containing two isolated charged
leptons with $p_T>15$~\GeV are rejected, to avoid overlap with the
dilepton channel. In addition, we require at least four jets
reconstructed using a midpoint cone algorithm~\cite{jetcone} with
radius ${\cal R}=0.5$, $p_T>20$~\GeV, and $|\eta|<2.5$; the jet with largest
transverse momentum must have $p_T > 40$~\GeV. 
Jets originating from $b$ quarks are identified using the output of a 
neural network where variables characterizing the
properties of secondary vertices and tracks with large impact
parameters relative to the \ppbar interaction vertex 
are combined~\cite{btagging}.



The \ttbar\ signal, with contributions from both $\qqbar \rightarrow \ttbar$ 
and $gg\rightarrow t\bar{t}$, is modeled using the \mcatnlo~\cite{mcatnlo}
event generator with the CTEQ6M1 parton distribution
functions (PDFs)~\cite{CTEQPDF}, assuming a top quark mass
$m_{t}=172.5$~\GeV.  
We generate \ttb\ MC samples both with and without the
expected spin correlation, corresponding to $A=0.78$ and $A=0$,
respectively~\cite{mcatnlo_LOtdecay}.
The events are further processed through \herwig~\cite{herwig}
to simulate parton evolution,  hadronization, and decays of
short-lived particles, followed by a full detector simulation 
using \geant~\cite{geant}. We overlay  events from random beam crossings
on the MC events  to model the effects of detector noise
and additional \ppbar\ interactions. The same reconstruction programs
are used to process the data and the simulated events. 

The background can be split into two components: multijet background,
where some of the products of hadronic partons are misreconstructed as
an isolated lepton, and inherent background from SM 
processes with  final states similar to that of the \ttbar signal. In
the \eplus\ channel, background from multijet production arises mainly when a
jet with high electromagnetic content mimics an electron. In the
\muplus channel it occurs primarily when a muon originates from the decay of a
heavy-flavor quark ($b$ or $c$) and appears to be isolated. The
multijet background is estimated from data~\cite{xsec_paper}.  
The SM background is predominantly from $W$+jets production,
with smaller contributions arising from single top quark, diboson
($WW$, $WZ$ and $ZZ$), and $Z$+jets ($Z\rightarrow ee$ in \ejets
or $Z\rightarrow \mu\mu$ in \mujets as well as $Z\rightarrow\tau\tau$)
events. The $W$+jets contribution is normalized to data
using an iterative procedure, where the expected \ttbar  and smaller
SM background contributions are subtracted from the data before
application of $b$-jet tagging~\cite{xsec_paper}. The differential
distributions for  $W$+jets are taken from a simulation using the
{\alpgen} MC program~\cite{alpgen}. 
All smaller SM background
contributions are also estimated using MC simulations but  
normalized to their next-to-leading-order predictions.
Diboson events are generated with {\pythia}~\cite{pythia}, single top quark
production with the {\textsc{comphep}} generator~\cite{CompHep}, and
$Z$+jets events are simulated using {\alpgen}.  
All MC background samples are generated using the CTEQ6L1
PDF~\cite{CTEQPDF}.  
The evolution of partons and the hadronization process are simulated 
using {\pythia}. A matching scheme is applied to avoid 
double-counting of partonic event configurations~\cite{matching}. 


To make optimal use of the kinematic information in \ttbar\ events, we
calculate signal probabilities $P_{\rm sgn}$ for each event using the
LO ME for the hypothesis of correlated ($H=c$) top quark 
spins, as predicted by the SM for $\qqbar \rightarrow \ttbar$,  and for the
hypothesis of uncorrelated ($H=u$) spins~\cite{Mahlon_Parke,nogg}. 
We can write $P_{\rm sgn}$ as a function of the hypotheses $H=c$ and $H=u$ as:  
\begin{eqnarray}
\label{eq:psgn}  \nonumber
P_{\rm sgn}(x;H) & & \hspace{-0.3cm}= \frac{1}{\sigma_{\rm obs}} \int
~f_{\rm PDF}(q_1)~f_{\rm PDF}(q_2) 
{\rm d}q_1 {\rm d}q_2 \\ 
&\times & \frac{(2 \pi)^{4} \left|{\cal M}(y,H)\right|^{2}}
       {q_1q_2 s}
  ~ W(x,y) 
~     {\rm d}\Phi_{6},
\end{eqnarray}
with $\sigma_{\rm obs}$ being the LO $\qqbar\rightarrow \ttbar$
production cross section including selection efficiency and acceptance
effects, $q_1$ and $q_2$ denoting the fraction of the proton and antiproton
momentum carried by the partons, 
$f_{\rm PDF}$ representing  the parton distribution functions, $s$ the
square of the center-of-mass energy of the colliding \ppbar system, 
and ${\rm d}\Phi_{6}$ the infinitesimal
volume element of the six-body phase space. Detector resolution effects 
are taken into account by introducing transfer functions $W(x,y)$
that describe the probability of a partonic final state $y$ to be
measured as $x = (\tilde{p}_1,\dots,\tilde{p}_n)$, where $\tilde{p}_i$
denote the measured four-momenta of the final state objects (leptons
and jets). 
For the  hypothesis $H=c$, we use the ME for the full process $q
\bar{q} \to \ttbar\ \to W^+ b \, W^- \bar{b} \to \ell^+ \nu_{\ell} b
\, q \bar{q}' \bar{b}$~\cite{chargeconjugation}
averaged over the color and spins of the initial partons, and summed over the
final colors and spins~\cite{Mahlon_Parke}. For the hypothesis $H=u$, we
use the ME for the same process, neglecting the spin correlation.  The
total \ttbar\ production cross section $\sigmatt$ and the selection
efficiency do not depend on spin correlation, thus the normalization
factor $\sigma_{\rm obs}$ can be omitted in Eq.~\ref{eq:psgn}. To
reduce the number of dimensions for the integrals, we assume the
directions of the momenta  of jets and charged leptons, and  
the electron energy are all well measured, and that the \ttbar\ system has
negligible transverse momentum. In addition, we use the known masses of the
final state particles as constraints. 

As we use only four jets when calculating $P_{\rm sgn}$, there are 24
possible jet-parton assignments. This further can be reduced to four
when identifying the jets originating from $b$ quarks. If more than
two jets are $b$-tagged, we select only the two jets with the largest $b$-tag
neural network probability as the $b$ jets, and assume other jets to
be light-flavor jets.  
Given the inability to distinguish the
flavor of the two quarks from the $W$ decay, as
required for the definition of the spin correlation
variable, both possible jet-parton assignments
have to be considered in the $P_{\rm sgn}$ calculation. 
Additional details of the  $P_{\rm sgn}$ calculation can be found in
Ref.~\cite{menimpaper}. 

To distinguish between correlated and uncorrelated top quark spin
hypotheses, we define, as in Ref.~\cite{d0_spin_me}, a 
 discriminant $R$~\cite{melnischulze},
\begin{equation}
R = \frac{P_{\rm sgn} (x;H=c)}{P_{\rm sgn}(x;H=u) + P_{\rm sgn} (x;H=c)}\,.
\label{eq:discr}
\end{equation}
To measure the ratio \fmeas of events with correlated spins to the total number of events,
we form templates from distributions of $R$ for \ttbar  MC events with and
without spin correlation as well as background. 
Since the main sources of background are from multijet and $W$+jets events, 
$P_{\rm sgn}$ is only calculated for these two contributions.  
The  smaller backgrounds are modeled using the templates for $W$+jets production.
The templates are compared to the distribution of $R$ in the data, and the
fraction \fmeas is extracted through a binned maximum-likelihood fit.   
To minimize the dependence of the result on  absolute normalization,
we calculate the predicted number of events as a function of \fmeas
and $\sigmatt$, and extract both simultaneously. Events used in the
templates are required to have at least two $b$-jet candidates; 
nonetheless, events with fewer than two $b$-tagged jets are
included in the fit to constrain the signal and background
normalization. The fitting procedure and $b$-jet identification
criteria are the same as used in Ref.~\cite{colorflow}. 

\begin{figure}[H,t]
\begin{centering}
\includegraphics[scale=0.45]{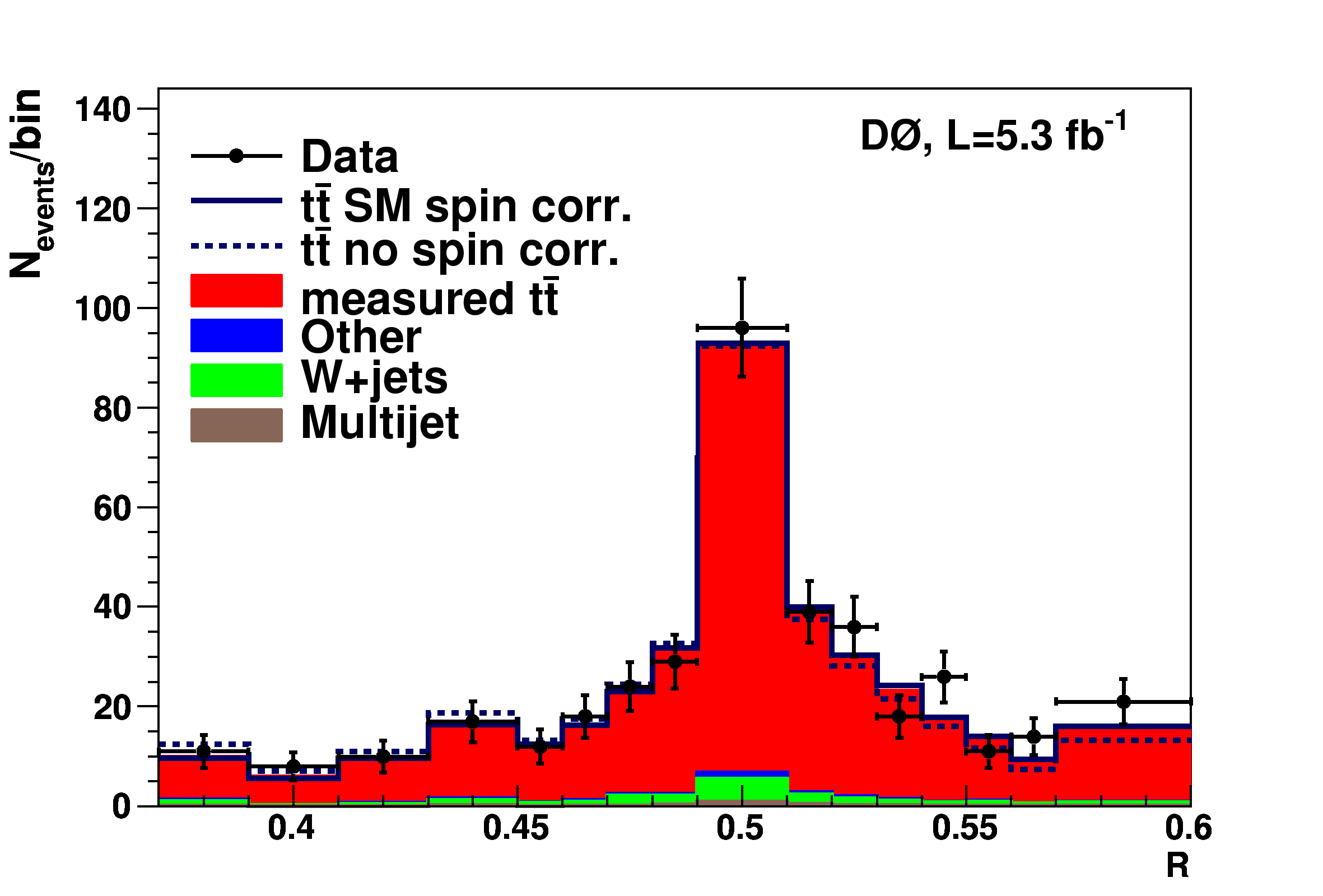}
\par\end{centering}
\caption{\label{fig:D_templates}
The distribution of the discriminant $R$ for \lplus events with four jets and 
an invariant mass of the two light-flavor jets within $\pm25~\GeV$ of the mass 
of the $W$ boson. 
The expectation (including background) for complete spin correlation as
predicted by the SM ($f=1$) and the case of no spin correlation
($f=0$), as well as the \ttbar contribution for \fmeas, where \fmeas was taken
from the combined fit in the \lplus and dilepton final states, are shown. The
first and last bins include contributions from $R<0.37$ and $R>0.60$,
respectively. The bin width is chosen to minimize the statistical uncertainty.}  
\end{figure}

To enhance the sensitivity, we divide events into four
subsamples as a correct jet-to-parton assignment greatly improves the
discrimination power of $R$. 
The events are divided into two groups of events with exactly four jets
and more than four jets to reduce the dilution from initial and final
state radiation. To reduce the contamination from events in which a
$b$-quark jet is mistakenly taken to come from a $W$ boson decay, 
these two groups are again separated according to whether the
invariant mass of the two light-flavor jets is within or outside of $\pm
25$~\GeV of the $W$ boson mass.
The $\pm 25$~\GeV window is based on optimization through pseudoexperiments.  
The main sensitivity to spin correlation is obtained in the subsample with four 
jets and a dijet invariant mass close to the $W$ boson mass, where the
probability of selecting the correct jet combination is the highest. 
In Fig.~\ref{fig:D_templates}, the measured discriminant $R$ for the
most sensitive sample is compared for data and templates of
\ttbar\ production with SM spin correlation and without spin
correlation, including background.  

\begin{figure}[H,b]
\begin{centering}
\includegraphics[scale=0.45]{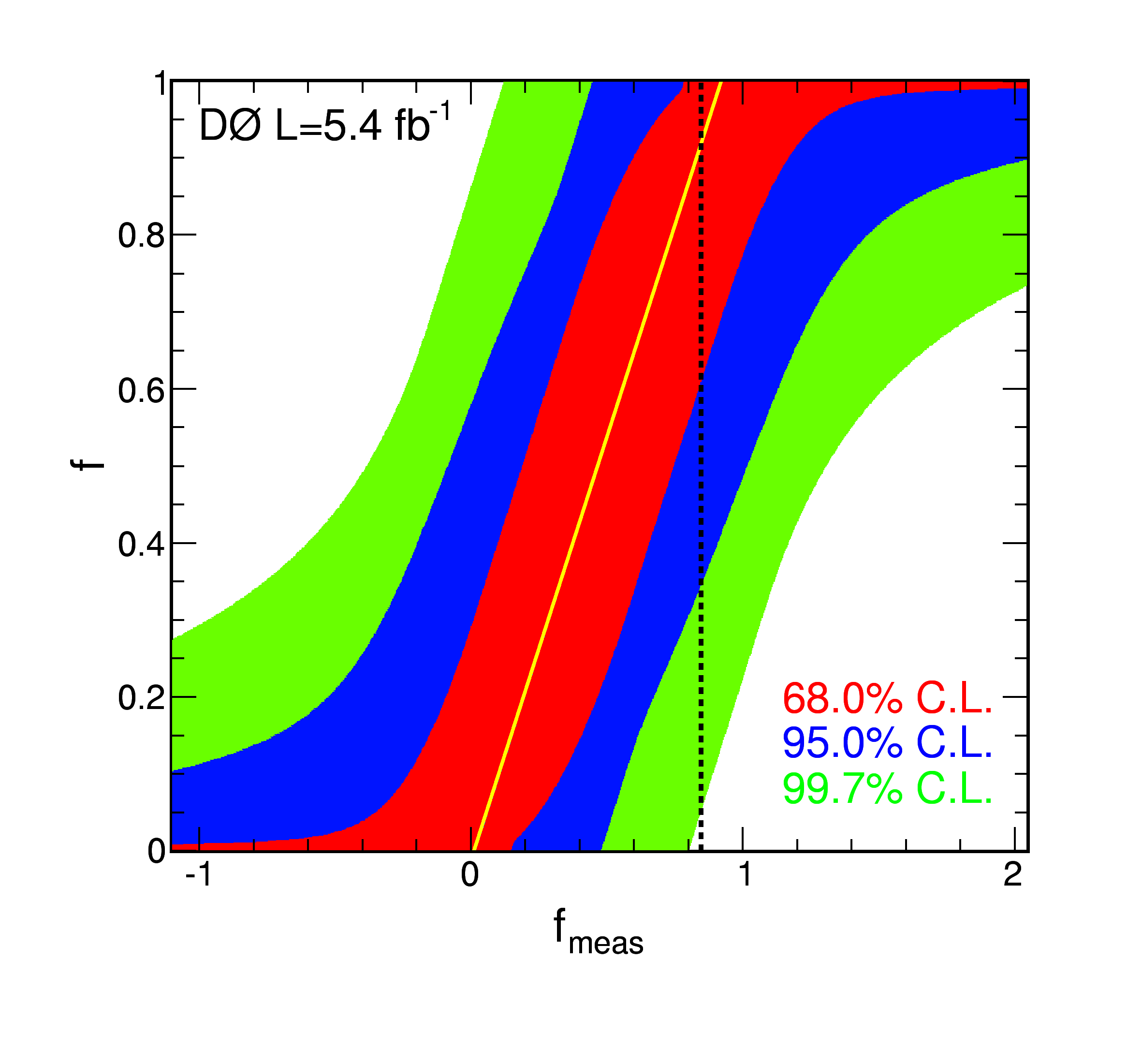}
\par\end{centering}
\caption{\label{fig:FC_combi}
(color online) Bands for  68\%, 95\% and 99.7\% C.L. of $f$ 
as a function of \fmeas for the combined dilepton and $\ell$+jets fit.
The thin light-color line indicates the most probable value of $f$ as a function of 
\fmeas. The vertical dotted black line shows the measured
value of $\fmeas=0.85$.   }
\end{figure}

We consider  the same systematic uncertainties as used in the
measurement of the \ttbar production cross section~\cite{xsec_paper} and \ttbar
spin correlation in dilepton events~\cite{d0_spin_me}. These  are
included in the likelihood fit through free parameters, where each
independent source of systematic uncertainty is modeled as a Gaussian
probability density function with zero mean and an rms corresponding
to one SD in the uncertainty on that parameter. Correlations among
systematic uncertainties for different channels are taken into account
by using a single parameter to represent the same source of uncertainty. 

We distinguish between systematic uncertainties that affect the 
yield of the signal or background and those that change  the distribution of $R$.  
We consider the jet energy scale, $b$-jet energy scale, jet energy
resolution, jet identification, $b$-tagging efficiency and $b$-jet
misidentification rate, choice of PDF, and the
choice of $m_t$ in the calculation of $P_{\rm sgn}$ as the
uncertainties that affect the distribution of $R$. 
Systematic uncertainties on normalizations include those on lepton
identification, trigger requirements, the normalization of background,
the luminosity, MC modeling, and the determination of multijet
background. We also include an uncertainty on the shape of the
templates varying each template bin within its statistical uncertainty. 
\begin{table}
\begin{center}
\caption{\label{tab:systematics_f} Summary of uncertainties on
  \fmeas for the combined fit in dilepton and \ljets channels. 
} 
\begin{tabular}{ccc} \hline \hline
 Source  &        +1SD    &   $-1$SD   \\ \hline
                       Muon identification &$ 0.003 $&$ -0.003 $\\ 
      Electron identification and smearing &$ 0.009 $&$ -0.008 $\\ 
                               PDF         &$ 0.058 $&$ -0.051 $\\ 
                               $m_t$       &$ 0.024 $&$ -0.040 $\\ 
                                 Triggers  &$ 0.007 $&$ -0.008 $\\ 
                 Opposite charge selection &$ 0.002 $&$ -0.002 $\\ 
                          Jet energy scale &$ 0.005 $&$ -0.028 $\\ 
     Jet reconstruction and identification &$ 0.007 $&$ -0.035 $\\ 
                               $b$-tagging &$ 0.012 $&$ -0.012 $\\ 
                            Normalization  &$ 0.039 $&$ -0.043 $\\ 
                            MC statistics  &$ 0.015 $&$ -0.015 $\\ 
                   Instrumental background &$ 0.003 $&$ -0.003 $\\ 
                                Luminosity &$ 0.023 $&$ -0.023 $\\ 
                       Multijet background &$ 0.007 $&$ -0.007 $\\ 
                                     Other &$ 0.007 $&$ -0.007 $\\ 
          MC statistics  for template fits &$ 0.156 $&$ -0.156 $\\ 
 \hline  
              Total systematic uncertainty &$ 0.176 $&$ -0.184 $\\ 
\hline  
                   Statistical uncertainty &$ 0.251 $&$ -0.258 $\\ 
\hline \hline
\end{tabular}
\end{center}
\end{table}

MC pseudoexperiments for different values of $f$ are used to estimate
the expected uncertainty on \fmeas, based on  the maximum-likelihood
fits that provide the dependence of $f$ on \fmeas. The ordering
principle for ratios of likelihoods~\cite{fc_limit} is applied to the
distributions of $f$ and \fmeas, without constraining \fmeas\ to the
physically allowed region. 
From a total of 729 events in the \lplus channels with a \ttbar signal
purity of 90\%, we obtain $\fmeas=1.15^{+0.42}_{-0.43}
{\rm(stat+syst)}$ and can exclude values of $f<0.420$ at the 95\% C.L.
Since the samples of dilepton~\cite{d0_spin_me} and $\ell$+jets final states
are statistically independent, results from the two channels can
be combined by adding the logarithms of the likelihood
functions and repeating the maximum-likelihood fit.   
We obtain 
\begin{equation}
\fmeas=0.85 \pm {0.29} {\rm (stat+syst)}
\end{equation}
and a \ttbar production cross section of
$\sigmatt=8.17^{+0.78}_{-0.67}~{\rm pb}$,  
which is in good agreement with the SM prediction~\cite{SMtheory_MochUwer} 
and previous measurements~\cite{xsec_paper}. The statistical and systematic 
uncertainties on \fmeas are given in Table~\ref{tab:systematics_f}.  
For an expected fraction of $f=1$, we can exclude $f<0.481$ at the
95\% C.L. For the observed value of $\fmeas=0.85$, we can exclude 
$f<0.344~(0.052)$ at the 95(99.7)\% C.L. 
We therefore obtain  first  evidence of SM spin correlation at 3.1 
standard deviations. The probability to have a true value
of $f=0$ for the observed value of $\fmeas=0.85$ is
$0.16$\%. Figure~\ref{fig:FC_combi} shows corresponding bands of
confidence level.  
The ratio \fmeas can be used to obtain a measurement
of the fractional difference \ameas by applying it as a multiplicative
factor to the NLO QCD prediction of \asm:~$\ameas=\fmeas\times\asm$.
This yields $\ameas=0.66\pm 0.23 \,({\text{stat+syst}})$~\cite{fintoc}. 

In conclusion, we have presented the first measurement of \ttbar spin
correlation using a matrix-element-based approach in the $\ell$+jets
channel. When combined with our previous result in the dilepton channel, we
obtain significant evidence for the presence of  spin correlation in \ttbar
events with 3.1 standard deviations.   
 
%
We thank the staffs at Fermilab and collaborating institutions,
and acknowledge support from the
DOE and NSF (USA);
CEA and CNRS/IN2P3 (France);
FASI, Rosatom and RFBR (Russia);
CNPq, FAPERJ, FAPESP and FUNDUNESP (Brazil);
DAE and DST (India);
Colciencias (Colombia);
CONACyT (Mexico);
KRF and KOSEF (Korea);
CONICET and UBACyT (Argentina);
FOM (The Netherlands);
STFC and the Royal Society (United Kingdom);
MSMT and GACR (Czech Republic);
CRC Program and NSERC (Canada);
BMBF and DFG (Germany);
SFI (Ireland);
The Swedish Research Council (Sweden);
and
CAS and CNSF (China).
%




\begin{thebibliography}{99}
\bibitem{eta}
  The pseudorapidity $\eta$ is defined relative to the center of the detector
as $\eta = -\ln [\tan(\theta/2)]$ where $\theta$ is the polar angle with
respect to the proton beam direction.
\bibitem{mcatnlo_LOtdecay}
In \mcatnlo the top quark decays are simulated
in LO QCD. We have checked that the difference between the spin
analyzing power $\alpha_{\bar{d}}$ in LO and NLO QCD does not impact
the final numerical result of \ameas.
\bibitem{nogg}
We do not use MEs for the smaller $gg\rightarrow t\bar{t}$
contribution in the calculation\ of $P_{\rm sgn}$, but account for it
in the measurement by using NLO \ttbar MC simulations with \qqbar and $gg$
production to extract $f$.
\bibitem{chargeconjugation}
Throughout this Letter, charge conjugated processes are included implicitly.
\bibitem{fintoc}
It should be noted that the allowed region for $f$ is restricted to
$0\leq f\leq 1$, resulting in the constraint
$0\leq A \leq 1$, while in principle $A$
can assume any value between -1 and 1. Therefore, values of $A$  
derived from $f$ can be compared to direct measurements of $A$ 
only when assuming that $A$ has a positive value, between 0 and the SM value.
\end{thebibliography}
\end{document}